\documentclass{revtex4}
\usepackage{amssymb}
\usepackage{amsmath}
\usepackage{graphicx}
\usepackage[font={footnotesize,it}]{caption}

\pdfoutput=1

\begin{document}

\title{Rindler Modified Schwarzschild Geodesics }
\author{M. Halilsoy}
\email{mustafa.halilsoy@emu.edu.tr}
\author{O. Gurtug}
\email{ozay.gurtug@emu.edu.tr}
\author{S. H. Mazharimousavi }
\email{habib.mazhari@emu.edu.tr}
\affiliation{Department of Physics, Eastern Mediterranean University, G. Magusa, north
Cyprus, Mersin 10, Turkey. }

\begin{abstract}
The mysterious attractive constant radial force acted in the past on Pioneer
spacecrafts - the so-called Pioneer anomaly - is considered within the
context of Rindler acceleration. As an idea this is tempting since it is
reminiscent of the cosmological constant. Fortunately the anomalous force
acts radially toward the sun so that it differs from the mission of a
cosmological constant. Without resorting to the physical source responsible
for such a term we investigate the modified Schwarzschild geodesics. The
Rindler acceleration naturally affects all massive / massless particle
orbits. Stable orbits may turn unstable and vice versa with a finely-tuned
acceleration parameter. The overall role of the extra term, given its
attractive feature is to provide confinement in the radial geodesics.
\end{abstract}

\maketitle

\section{INTRODUCTION}

In an attempt to describe gravity of a central mass at large distances and
explain certain satellite anomalies Grumiller \cite{1} gave a new metric in
static spherically symmetric (SSS) spacetimes. The anomaly refers to the
test particle behavior in the field of pairs, such as Sun-Pioneer
spacecraft, Earth-satellites etc. and therefore it is of utmost importance 
\cite{2}. The ansatz of SSS, naturally reduces the problem effectively to a
two-dimensional system where the Newtonian central force modifies into $%
Force=-(\frac{M}{r^{2}}+a)$. Here $M$ is the central gravitating mass while $%
a=const.$ is known as the Rindler acceleration parameter. For $a>0$, the two
forces are both toward center while $a<0,$ gives an outward repulsive force.
Throughout our analysis in this paper we shall be using $a>0$, unless stated
otherwise. In particular, the attractive mysterious force acting on the
Pioneer spacecrafts (10/11) launched in 1972 / 73 which could not be
accounted by any known physical law came to be known as the Pioneer anomaly.
We must add that a recent proposal based on thermal heat loss of the
satellite may resolve the anomaly \cite{ANOMALY}. Among other proposals
Modified Newtonian Dynamics (MOND) in the outer space attracted also
interest. The fact that the Rindler force is a constant one dashes hopes to
treat it as a perturbation due to other sources (or planets). Further, the
non-isotropic role of parameter $a$ differs it from the cosmological
constant. Although it remains still open to investigate about the physical
source that gives rise to such a term, herein we wish to study geodesics in
the resulting spacetime. As a matter of fact in a separate study we have
shown consistently that the Rindler acceleration parameter can be attributed
to non-linear electrodynamics (NED) \cite{3} ([4a]) and to a space-filling
fluid in $f(R)$ gravity \cite{3} ([4b]). In such a formalism the weak (and
strong) energy conditions are satisfied. Such nice features, however, are
not without pay off: global monopoles emerge as complementary objects to the
energy conditions \cite{3} ([4a]). For vanishing Rindler acceleration, $a=0$
, all results reduce to those of Schwarzschild as it should. Experimental
estimates suggest that for the satellites $a\sim 10^{-10}m/s^{2}$, which
suffices to yield significant differences at large distances. Let us add
that in the vicinity of a Schwarzschild horizon $r=2m,$ any object feels a
constant acceleration toward the black hole a l\'{a} Rindler. However, as
stated above the Rindler acceleration in question refers to large distances
and for this the central object need not be a black hole. In particular such
a constant acceleration at certain distances from Earth (or any other
planet) effects satellite motion to the extent that are dubbed as
"anomalies". It should be, added that, in order that such 'anomalies' are
universal each satellite / planet must feel it equally. So far no such
anomaly is reported in objects other than the Pioneer spacecrafts.

The physical effects of the model proposed by Grumiller have been
investigated in \cite{4}, by studying the classical solar system tests of
general relativity. In that study, perihelion shifts, light bending and
gravitational redshift were calculated for solar system planets in the
presence of Rindler parameter. In another study \cite{5}, an alternative
method has been employed to calculate the bending of light that affects the
bounds on the Rindler parameter.

The main motivation of the present paper is to investigate the effect of the
Rindler parameter $a$, on the trajectories of the timelike and null
geodesics. Our analysis shows that the Rindler acceleration tends to confine
the geodesics. It would not be wrong to state that, with this remarkable
property finely-tuned, Rindler parameter serves to have a kind of bounded
cosmology.

No doubts the best account of Schwarzschild geodesics is given in "The
Mathematical Theory of Black Holes" by Chandrasekhar \cite{6}. Being
integrable systems the geodesic equations do not exhibit chaotic behavior.
It is known that in some gravitational systems such as Kundt Type-III and
type-N spacetimes chaos is indispensable \cite{7} (and references cited
therein). The Rindler modified Schwarzschild spacetime on the other hand is
a type-D spacetime.

In this paper, we follow a similar procedure as was done for Schwarzschild
metric \cite{6}. To be more precise, we will closely follow the method used
in \cite{8}. The motivation behind this is to compare the effect of Rindler
parameter that deviates the results from the analysis performed in \cite{6}.
The paper is organized as follows. The review of Grumiller's spacetime and
its type-D character is emphasized in section II. In section III, the
derivation of the particles trajectorie for timelike and null geodesics with
their numerical plots are presented. Null geodesics in terms of elliptic
functions are given in the Appendix. Our Conclusion is given in section IV.

\section{STRUCTURE OF GRUMILLER'S SPACETIME}

\begin{figure}[tbp]
\includegraphics[width=90mm,scale=0.9]{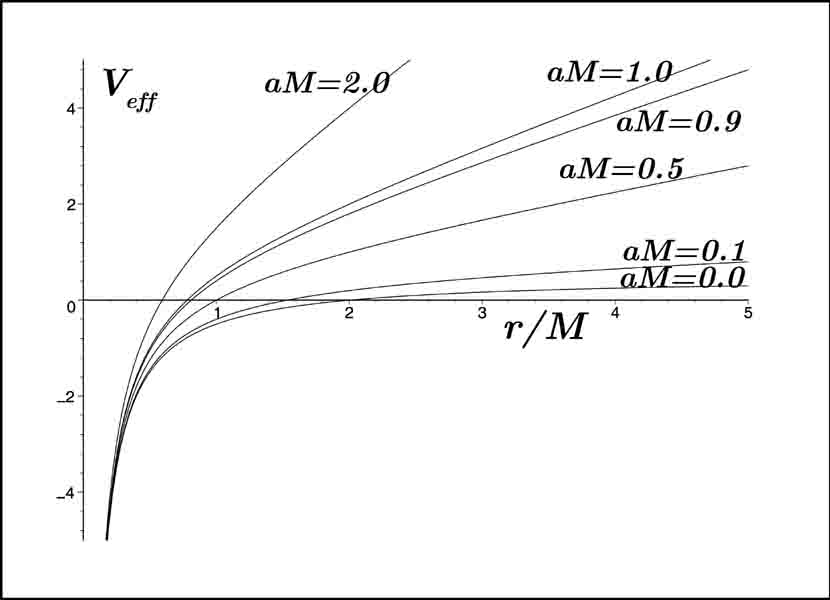}
\caption{The effective potential $V_{eff}$ (Eq. 26) rises with the
increasing $a$. This behavior serves to confine geodesics nearer to the
gravitating center. }
\end{figure}

Grumiller have proposed a model for gravity at large distances of a central
object by assuming static and spherically symmetric system. The
corresponding line element in the absence of cosmological constant is given
by%
\begin{equation}
ds^{2}=-f\left( r\right) dt^{2}+\frac{dr^{2}}{f\left( r\right) }+r^{2}\left(
d\theta ^{2}+\sin ^{2}\theta d\phi ^{2}\right) ,
\end{equation}%
where%
\begin{equation}
f\left( r\right) =1-\frac{2M}{r}+2ar.
\end{equation}%
Herein $M$ is the mass of the central object (or black hole) and $a\geq 0$
is a real constant. With $a=0,$ Eq. (1) reduces to the Schwarzschild black
hole. The only horizon of the metric is given by $f\left( r\right) =0$ which
yields%
\begin{equation}
r_{h}=\frac{-1+\sqrt{1+16Ma}}{4a}.
\end{equation}%
It can easily be checked that for $a\rightarrow 0,$ we obtain $r_{h}=2M$, as
it should be. Furthermore the scalars of the spacetime are given by%
\begin{eqnarray}
R &=&-12\frac{a}{r}, \\
R_{\mu \nu }R^{\mu \nu } &=&40\frac{a^{2}}{r^{2}}, \\
K &=&R_{\mu \nu \alpha \beta }R^{\mu \nu \alpha \beta }=48\frac{M^{2}}{r^{6}}%
+32\frac{a^{2}}{r^{2}}
\end{eqnarray}%
which indicate the typical central curvature singularity at $r=0$ behind the
event horizon.

\subsection{The Description of the Grumiller Spacetime in a Newman-Penrose
(NP) Formalism}

\begin{figure}[tbp]
\includegraphics[width=180mm,scale=1]{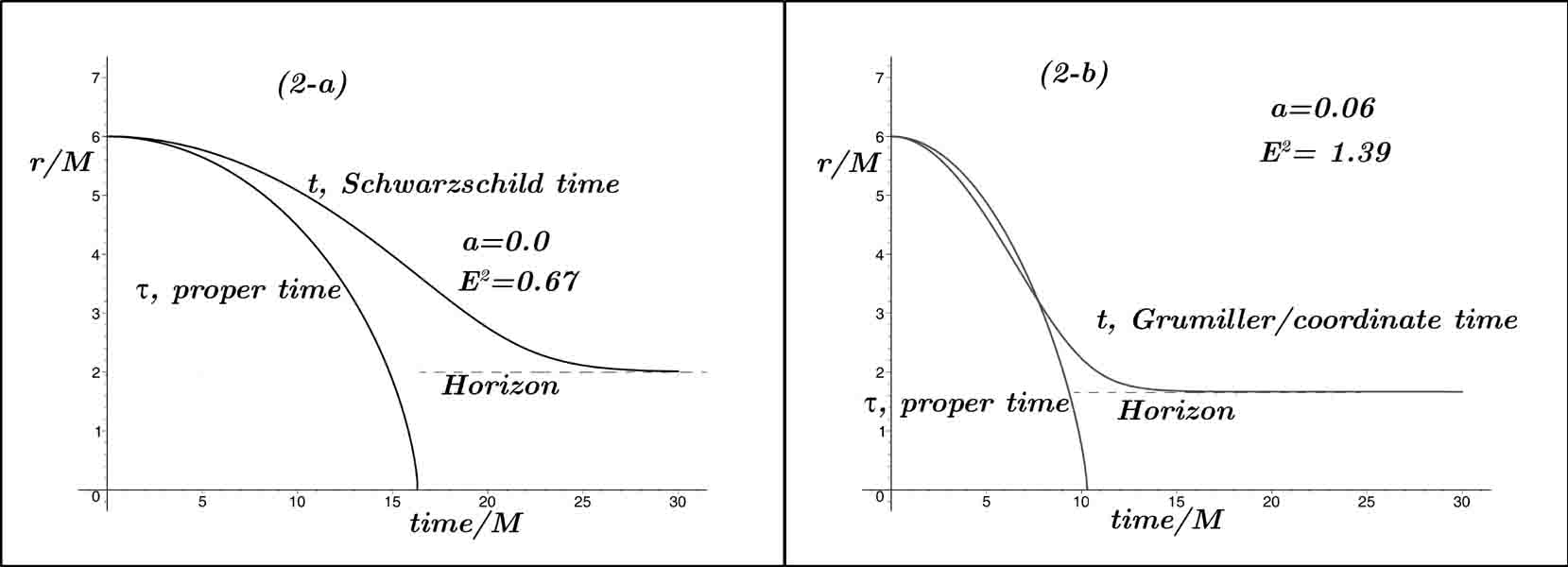}
\caption{Particle-fall into $r=0$ singularity versus times (i.e. coordinate
and proper) in (a) Schwarzschild and (b) Grumiller metric. Outside the
horizon criss-crossing of geodesics in (b) differs from the Schwarzschild
case. The fall evidently delays in (a) compared to (b).}
\end{figure}

The Grumiller's metric is investigated with the Newman-Penrose (NP)
formalism \cite{9} in order to explore the physical properties of the metric
or clarifying the role of Rindler parameter on the physical quantities. Note
that the signature of the metric given in (1) is changed in this section to $%
(+,-,-,-)$, apt for the NP formalism. \ The set of proper null tetrad $1-$%
forms is given by%
\begin{eqnarray}
l &=&dt-\frac{dr}{f(r)}, \\
n &=&\frac{1}{2}\left( f(r)dt+dr\right) ,  \notag \\
m &=&-\frac{r}{\sqrt{2}}\left( d\theta +i\sin \theta d\varphi \right) , 
\notag
\end{eqnarray}%
and the complex conjugate of $m.$ The non-zero spin coefficients in this
tetrad are%
\begin{equation}
\beta =-\alpha =\frac{\cot \theta }{2\sqrt{2}r},\text{ \ \ }\rho =-\frac{1}{r%
},\text{ \ \ }\mu =-\frac{f(r)}{2r},\text{ \ \ }\gamma =\frac{1}{4}\frac{%
df(r)}{dr}.
\end{equation}%
We obtain as a result, the Weyl and Ricci scalars as%
\begin{eqnarray}
\Psi _{2} &=&-\frac{M}{r^{3}}, \\
\phi _{11} &=&\Lambda =-\frac{a}{2r},  \notag
\end{eqnarray}%
so that the spacetime is Petrov type-D. The Rindler acceleration parameter
is seen to effect only the Ricci components, leaving the mass term $\Psi
_{2} $ of Schwarzschild unchanged. In the so called Pioneer anomaly the
attractive force suggests that $a>0$, which is the case that we shall
investigate in this paper. However, the fact that $\phi _{11}<0,$ under the
choice $a>0$ which violates the energy conditions suggests also that the
underlying physical source is not a familiar one. In a separate study \cite%
{3}, we showed explicitly what the physical source may be, and we reached
the conclusion that the arrow \ indicates an unaccustomed source: the
non-linear electrodynamics (NED). Pure electric/magnetic sources with
unusual Lagrangian of NED does yield the Rindler-type acceleration term
without violating the energy conditions. We restate that, in this paper, we
shall investigate only the resulting geodesics without touching the possible
physical source of the Rindler parameter $a$.

\section{PARTICLE TRAJECTORIES}

\begin{figure}[tbp]
\includegraphics[width=90mm,scale=0.9]{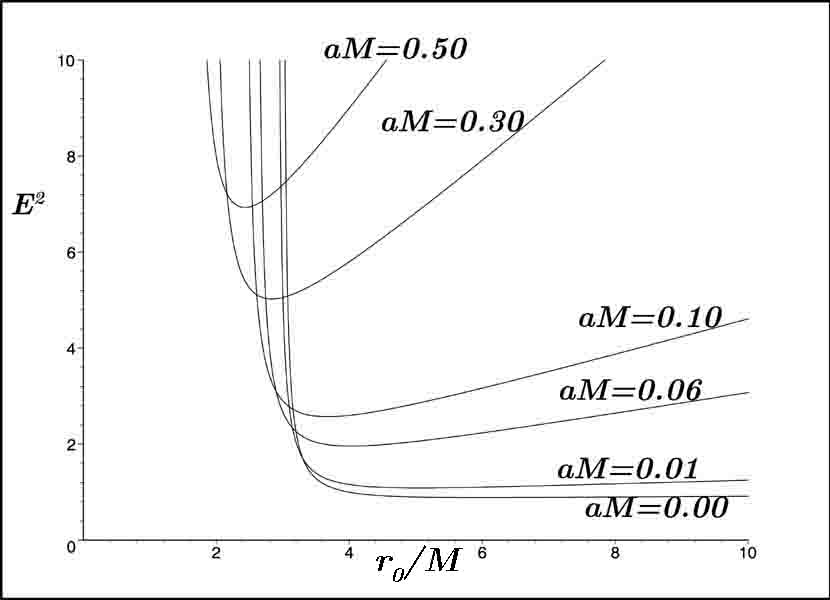}
\caption{A plot of $E^{2}$ versus $r_{0}/M$ (Eq. (32)) of a massive particle
starting at rest from $r_{0}$, moving along a timelike geodesic for various
values of $aM$.}
\end{figure}

The Lagrangian for a massive particle with unit mass is given by%
\begin{equation}
\mathcal{L}=\frac{1}{2}g_{\mu \nu }\dot{x}^{\mu }\dot{x}^{\nu }=-\frac{1}{2}%
f\left( r\right) \dot{t}^{2}+\frac{\dot{r}^{2}}{2f\left( r\right) }+\frac{1}{%
2}r^{2}\left( \dot{\theta}^{2}+\sin ^{2}\theta \dot{\phi}^{2}\right)
\end{equation}%
in which a dot "$^{\cdot }$" shows derivative with respect to an arbitrary
parameter $\sigma $. There are two conserved quantities, the energy $E$ and
the angular momentum $\ell $ in $\phi $ direction,%
\begin{equation}
E=\frac{\partial \mathcal{L}}{\partial \dot{t}}=g_{tt}\frac{dt}{d\sigma }%
=-f\left( r\right) \frac{dt}{d\sigma }=\text{constant}
\end{equation}%
and 
\begin{equation}
\frac{\partial \mathcal{L}}{\partial \dot{\phi}}=r^{2}\sin ^{2}\theta \dot{%
\phi}=\ell =\text{constant}.
\end{equation}%
Note that $\ell =r^{2}\dot{\phi}$ is the angular momentum of the particle
moving on the plane $\theta =\frac{\pi }{2}.$ For null ($\epsilon =0$) and
timelike ($\epsilon =1$) geodesics one has, 
\begin{equation}
g_{\mu \nu }\frac{dx^{\mu }}{d\sigma }\frac{dx^{\nu }}{d\sigma }=-\epsilon
\end{equation}%
which on the plane of motion $\theta =\frac{\pi }{2}$ implies%
\begin{equation}
\left( \frac{dr}{d\sigma }\right) ^{2}=E^{2}-f\left( r\right) \left(
\epsilon +\frac{\ell ^{2}}{r^{2}}\right) .
\end{equation}%
This equation can be cast into a familiar form of equation of motion for a
unit mass test particle 
\begin{equation}
\frac{1}{2}\left( \frac{dr}{d\sigma }\right) ^{2}+V_{eff}\left( r\right) =%
\mathcal{E}_{eff}
\end{equation}%
with an effective potential $V_{eff}\left( r\right) =\frac{1}{2}f\left(
r\right) \left( \epsilon +\frac{\ell ^{2}}{r^{2}}\right) $ and corresponding
effective energy $\mathcal{E}_{eff}=\frac{1}{2}E^{2}.$ The exact form of the
effective potential can be written as%
\begin{equation}
V_{eff}\left( r\right) =\frac{1}{2}\left( 1-\frac{2M}{r}+2ar\right) \left(
\epsilon +\frac{\ell ^{2}}{r^{2}}\right) .
\end{equation}%
We note that the classical region which $r$ may take is limited by the
constraint $\mathcal{E}_{eff}\geq V_{eff}\left( r\right) $ to keep $r$ real.
Using the chain rule $\frac{dr}{d\sigma }=\frac{dr}{d\phi }\frac{\ell }{r^{2}%
},$ one finds%
\begin{equation}
\left( \frac{dr}{d\phi }\right) ^{2}=\frac{2r^{4}}{\ell ^{2}}\left( \mathcal{%
E}_{eff}-V_{eff}\left( r\right) \right) .
\end{equation}%
As in the standard Kepler problem for convenience, we introduce $r=\frac{1}{u%
}$ which yields the $u-$equation for future use%
\begin{equation}
\left( \frac{du}{d\phi }\right) ^{2}=\frac{E^{2}}{\ell ^{2}}-\left( 1-2Mu+%
\frac{2a}{u}\right) \left( \frac{\epsilon }{\ell ^{2}}+u^{2}\right) .
\end{equation}%
In the sequel we shall use this general equation to investigate the motion
of the particle in different cases.

\subsection{Motion with zero angular momentum: radial}

\begin{figure}[tbp]
\includegraphics[width=90mm,scale=0.9]{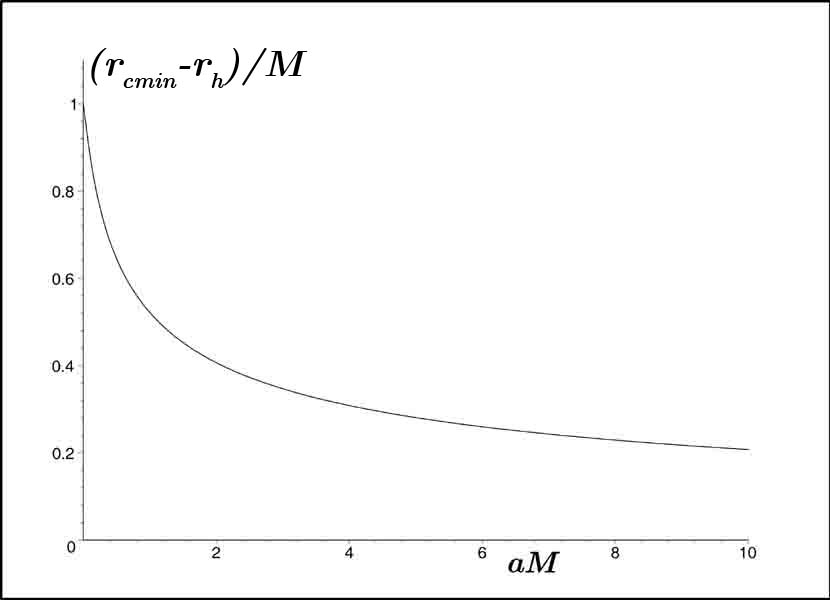}
\caption{The difference between circular geodesics radii ($r_{c\min }$) and
horizon radii ($r_{h}$) is shown as a function of $aM$. For increasing $aM$
it shows that the circular geodesics approach to the horizon.}
\end{figure}

In zero angular momentum case (i.e. $\ell =0$) the motion will remain in the
plane $\phi =$constant and the particle will move radially. This in turn
implies from (14) that%
\begin{equation}
\left( \frac{dr}{d\sigma }\right) ^{2}=E^{2}-\epsilon f\left( r\right) .
\end{equation}%
In the following subsections we shall study the cases $\epsilon =0$ (null)
and $\epsilon =1$ (timelike) separately. First, let's consider the null
geodesics (i.e., $\epsilon =0$):

\subsubsection{Null geodesics $\protect\epsilon =0$}

\begin{figure}[tbp]
\includegraphics[width=180mm,scale=1]{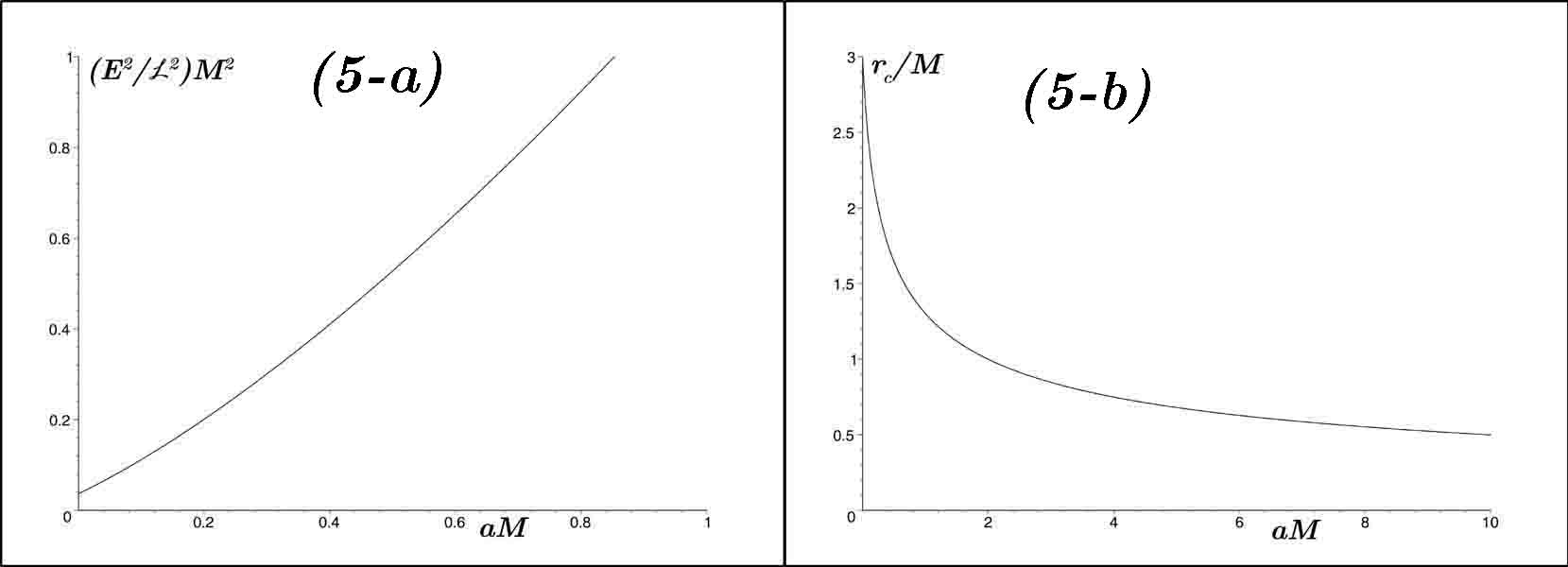}
\caption{ Plot of $\left( \frac{E^{2}}{\ell ^{2}}\right) M^{2}$ and $\frac{%
r_{c}}{M}$ (Eq. (41) and (40)) in terms of $aM$ for a massless particle.
These figures show that for larger $a$ (with a given mass $M$) the circular
orbit of the photon has smaller radius but larger value of $\left( \frac{%
E^{2}}{\ell ^{2}}\right) .$}
\end{figure}

In null geodesics, which refers to the motion of a massless particle
(photon), Eq. (19) becomes%
\begin{equation}
\left( \frac{dr}{d\sigma }\right) ^{2}=E^{2}.
\end{equation}%
We recall from (11) that $E=-f\left( r\right) \frac{dt}{d\sigma }$ and
therefore after some modification this equation becomes%
\begin{equation}
\frac{dr}{dt}=\pm f\left( r\right) =\pm \left( 1-\frac{2M}{r}+2ar\right) ,
\end{equation}%
which explicitly admits the following integral between the time and the
radial position of the photon,%
\begin{equation}
\frac{\ln \left( \frac{r-2M+2ar^{2}}{r_{0}-2M+2ar_{0}^{2}}\right) }{4a}+%
\frac{\left[ \tanh ^{-1}\left( \frac{1+4ar}{\sqrt{16Ma+1}}\right) -\tanh
^{-1}\left( \frac{1+4ar_{0}}{\sqrt{16Ma+1}}\right) \right] }{2a\sqrt{16Ma+1}}%
=\pm \left( t-t_{0}\right) .
\end{equation}%
Herein $r_{0}$ is the initial position of the massless particle (photon) and 
$t$ is the time measured by the distant observer while $t_{0}$ is the
initial time.

\subsubsection{Timelike Geodesics $\protect\epsilon =1$}

\begin{figure}[tbp]
\includegraphics[width=90mm,scale=0.9]{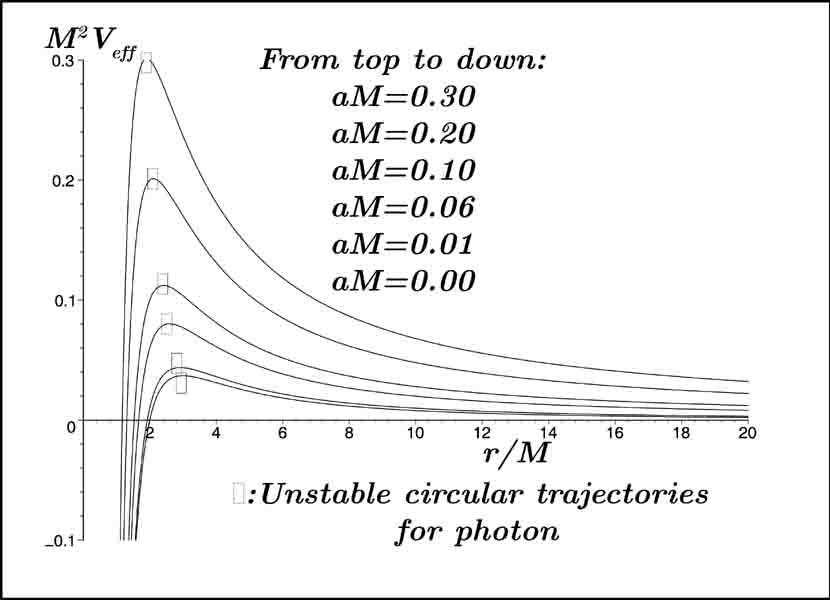}
\caption{The $M^{2}V_{eff}$ plot versus $\frac{r}{M}$ for photons shows the
unstable circular orbits for photons. In this range of $aM$ no stable photon
orbits exist.}
\end{figure}

In the case of timelike geodesics (i.e., $\epsilon =1$), which refers to the
motion of a massive particle, with unit mass Eq. (19) becomes 
\begin{equation}
\left( \frac{dr}{d\sigma }\right) ^{2}=E^{2}-1+\frac{2M}{r}-2ar,
\end{equation}%
which can be cast into the following equation of motion 
\begin{equation}
\frac{d^{2}r}{d\sigma ^{2}}=-\frac{M}{r^{2}}-a.
\end{equation}%
Choosing $\sigma $ to be the proper time, $\tau $, one finds 
\begin{equation}
\frac{d^{2}r}{d\tau ^{2}}=-\left( \frac{M}{r^{2}}+a\right) .
\end{equation}%
Considering $a>0,$ clearly the radial force per unit mass is attractive and
toward the central object / black hole. Now let's consider the particle
initially at rest and, upon the gravitational attraction, starts moving from
its initial radial location $r=r_{0}$. Using (23) with $\sigma =\tau $ one
finds%
\begin{equation}
E^{2}=1-\frac{2M}{r_{0}}+2ar_{0}
\end{equation}%
and therefore%
\begin{equation}
\left( \frac{dr}{d\tau }\right) ^{2}=2a\left( r_{0}-r\right) +2M\left( \frac{%
1}{r}-\frac{1}{r_{0}}\right) .
\end{equation}%
Further, the effective potential in radial motion reads 
\begin{equation}
V_{eff}\left( r\right) =\frac{1}{2}\left( 1-\frac{2M}{r}+2ar\right)
\end{equation}%
and upon introducing $\frac{r}{M}=\tilde{r},$ $Ma=\tilde{a}$ we find $%
V_{eff}\left( \tilde{r}\right) =\frac{1}{2}\left( 1-\frac{2}{\tilde{r}}+2%
\tilde{a}\tilde{r}\right) ,$ which is depicted in Fig. 1. In this figure the
horizon $\tilde{r}_{h}$ is the intersection of $V_{eff}\left( \tilde{r}%
\right) $ with $\tilde{r}$ axis. This figure shows that with the Rindler
parameter there is an upper bound for the motion of the particle. Hence,
unlike the Schwarzschild spacetime with $\mathcal{E}_{eff}\geq 1$(for $a=0$
/Schwarzschild spacetime, there is an upper bound for the motion of the
particle if $\mathcal{E}_{eff}\leq 1$.), the particle can not escape to
infinity. Finally, using Eq.s (23) and (11) one finds%
\begin{equation}
\left( \frac{dr}{dt}\right) ^{2}=\frac{1}{E^{2}}\left( E^{2}-1+\frac{2M}{r}%
-2ar\right) \left( 1-\frac{2M}{r}+2ar\right) ^{2}
\end{equation}%
which after differentiating with respect to $t$ one obtains,%
\begin{equation}
\frac{d^{2}r}{dt^{2}}=-\frac{12}{E^{2}r^{4}}\left( M+ar^{2}\right) \left(
M-ar^{2}-\frac{r}{2}\right) \left( M+\left( \frac{E^{2}}{3}-\frac{1}{2}%
\right) r-ar^{2}\right) .
\end{equation}%
Here also $t$ is the time measured by the distant observer. In Fig.s 2a and
2b we plot the variation of the coordinate time ($t$) and the proper time ($%
\tau $) along a timelike radial-geodesic described by the test particle,
starting at rest at $r=r_{0}$ and falling towards the singularity. We
comment that the conserved energy of the particle is found from Eq. (29),
i.e. 
\begin{equation}
0=\frac{1}{E^{2}}\left( E^{2}-1+\frac{2M}{r_{0}}-2ar_{0}\right) \left( 1-%
\frac{2M}{r_{0}}+2ar_{0}\right) ^{2}
\end{equation}%
and consequently%
\begin{equation}
E^{2}=1-\frac{2M}{r_{0}}+2ar_{0}.
\end{equation}%
From Fig.s 2a and 2b, it is observed that the Rindler parameter accelerates
the particles following the timelike geodesics so that the particles reach
the singularity faster than the Schwarzschild case. In Fig. 3 we plot $E^{2}$
versus $\frac{r_{0}}{M}$ for different values of $aM.$ This figure clearly
shows that for non-zero $a,$ energy possesses a minimum value. 
\begin{figure}[tbp]
\includegraphics[width=90mm,scale=0.7]{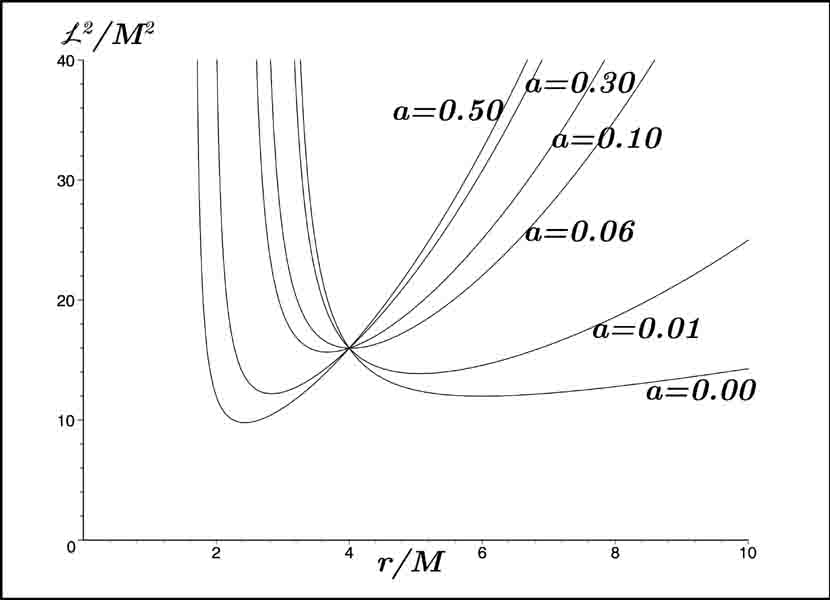}
\caption{Angular momentum behavior of a massive particle versus distance for
changing $a$ (Eq. (43)). It is observed that all curves coincide at $\frac{r%
}{M}=4.$}
\end{figure}

\subsection{Circular Motion}

After considering the radial motion, in this section we study the circular
motion of a photon and a test massive particle by considering $\left. \frac{%
du}{d\phi }\right\vert _{u=u_{c}}=0$ (Eq. (18)) in which $r_{c}=1/u_{c}$ is
the circular orbit of the particle. This condition, in turn implies 
\begin{equation}
\frac{E^{2}}{\ell ^{2}}-\left( 1-2Mu_{c}+\frac{2a}{u_{c}}\right) \left( 
\frac{\epsilon }{\ell ^{2}}+u_{c}^{2}\right) =0.
\end{equation}%
Once more let's look at Eq. (18) and find derivative of both sides with
respect to $\varphi $ which gives%
\begin{equation*}
\frac{d^{2}u}{d\phi ^{2}}=\frac{d}{du}\left( \frac{E^{2}}{\ell ^{2}}-\left(
1-2Mu+\frac{2a}{u}\right) \left( \frac{\epsilon }{\ell ^{2}}+u^{2}\right)
\right) .
\end{equation*}%
In order to have equilibrium motion $\frac{d^{2}u}{d\phi ^{2}}=0$ must hold
and this in turn implies 
\begin{equation}
\left. \frac{d}{du}\left[ \frac{E^{2}}{\ell ^{2}}-\left( 1-2Mu+\frac{2a}{u}%
\right) \left( \frac{\epsilon }{\ell ^{2}}+u^{2}\right) \right] \right\vert
_{u=u_{c}}=0.
\end{equation}%
These conditions yield the expressions for the angular momentum $\ell $ and
the energy $E$ of the particle as%
\begin{equation}
\ell ^{2}=\frac{\epsilon \left( Mu_{c}^{2}+a\right) }{u_{c}^{2}\left(
a+u_{c}-3Mu_{c}^{2}\right) },
\end{equation}%
\begin{equation}
E^{2}=\frac{4\epsilon \left( a+\frac{u_{c}}{2}-Mu_{c}^{2}\right) ^{2}}{%
u\left( a+u_{c}-3Mu_{c}^{2}\right) }.
\end{equation}%
We note that the considered spacetime is nonasymptotically flat. Hence, the
value of $r_{c}$ is bounded. For a physically acceptable motion the
constraint $a+u_{c}-3Mu_{c}^{2}>0$ arises naturally from Eq. (35) which in
turn admits $r_{c}>\frac{-1+\sqrt{1+12Ma}}{2a}=r_{c\min }.$ Here $r_{c\min }$
is clearly larger than the horizon $r_{h}=\frac{-1+\sqrt{1+16Ma}}{4a}$ so
that $r_{c\min }-r_{h}=\frac{2\sqrt{1+12Ma}-\sqrt{1+16Ma}-1}{4a}>0.$ In Fig.
4 we plot $\frac{r_{c}-r_{h}}{M}$ in terms of $Ma$ which implies that with
larger $a,$ although the gap between $r_{c\min }$ and $r_{h}$ gets smaller,
it always remains positive. Note that to have a physical orbit $r_{c}>r_{h}$
must hold.

\begin{figure}[tbp]
\includegraphics[width=180mm,scale=1]{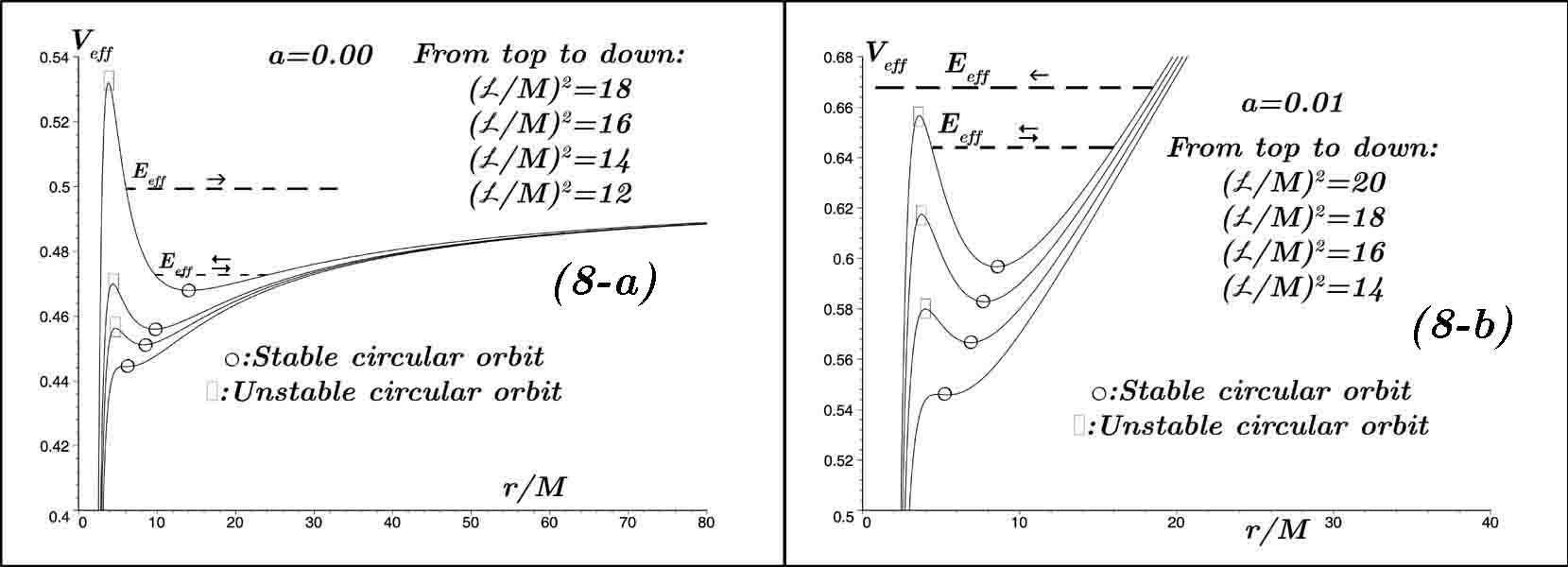}
\caption{$V_{eff}$ for massive particles versus $\frac{r}{M}$ for (a) the
Schwarzschild case ($a=0$) and (b) the Grumiller case with $a=0.01$ and
various values of $\frac{E^{2}}{\ell ^{2}}$. The potential barriers are
comparatively shown in both cases. The steeply rising potential barrier in
(8b) explains why the orbits remain bounded. This doesn't occur in (8a). The
minimum stable orbit can be obtained from $\left. \frac{dV_{eff}}{dr}%
\right\vert _{r=r_{\min }}=0$ and $\left. \frac{d^{2}V_{eff}}{dr^{2}}%
\right\vert _{r=r_{\min }}=0$ , which yields the Rindler parameter $a$ in
terms of $r_{\min }$ as $a=\frac{\ell ^{2}}{r_{\min }^{3}}\left( \frac{3M}{%
r_{\min }}-\frac{1}{2}\right) .$}
\end{figure}

\subsubsection{Null geodesics ($\protect\epsilon =0$)}

In the case of null geodesics we set $\epsilon =0$ in Eq.s (33) and (34) to
obtain 
\begin{equation}
r_{c}=\frac{-1+\sqrt{1+12aM}}{2a}
\end{equation}%
and 
\begin{equation}
\frac{E^{2}}{\ell ^{2}}=\frac{\left( 1+\sqrt{1+12aM}+24aM\right) \left( 1+%
\sqrt{1+12aM}\right) }{108M^{2}}.
\end{equation}%
Having $r_{c}$ known an exact value for the specific $M$ and $a$ means that
for the photon there exist only one equilibrium circular orbit with the
ratio $\frac{E^{2}}{\ell ^{2}}$ given in the latter equation. In Fig. 5 we
plot $\frac{E^{2}M^{2}}{\ell ^{2}}$ and $\frac{r_{c}}{M}$ versus $aM.$ It
shows that for larger $a$ (with a given mass $M$) the circular orbit of the
photon has larger value of $\left( \frac{E^{2}}{\ell ^{2}}\right) $ and
smaller value of $r_{c}.$

To complete present analysis we study the stability of such orbits. This can
be done by considering the geodesic equation of the photon in Eq. (14) with $%
\epsilon =0$ which becomes 
\begin{equation}
\frac{1}{\ell ^{2}}\left( \frac{dr}{d\sigma }\right) ^{2}+\frac{f\left(
r\right) }{r^{2}}=\frac{E^{2}}{\ell ^{2}}.
\end{equation}%
A replacement of $\sigma =$ $\frac{\tilde{\sigma}}{\ell }$ gives%
\begin{equation}
\left( \frac{dr}{d\tilde{\sigma}}\right) ^{2}+V_{eff}=E_{eff}
\end{equation}%
in which 
\begin{equation}
V_{eff}=\frac{f\left( r\right) }{r^{2}}=\frac{1-\frac{2M}{r}+2ar}{r^{2}}
\end{equation}%
and 
\begin{equation}
E_{eff}=\frac{E^{2}}{\ell ^{2}}.
\end{equation}%
To have a stable circular orbit we must have $\left. V_{eff}^{\prime
}\right\vert _{r_{c}}=0$ and $\left. V_{eff}^{\prime \prime }\right\vert
_{r_{c}}>0.$ We have plotted $V_{eff}$ in Fig. 6, showing that, there is no
stable circular orbit for photons. As a remark to this section let us add
that in case that the source of the Rindler acceleration is NED \cite{4} the
null geodesics do not represent in general the photon orbits. 
\begin{figure}[tbp]
\includegraphics[width=160mm,scale=0.9]{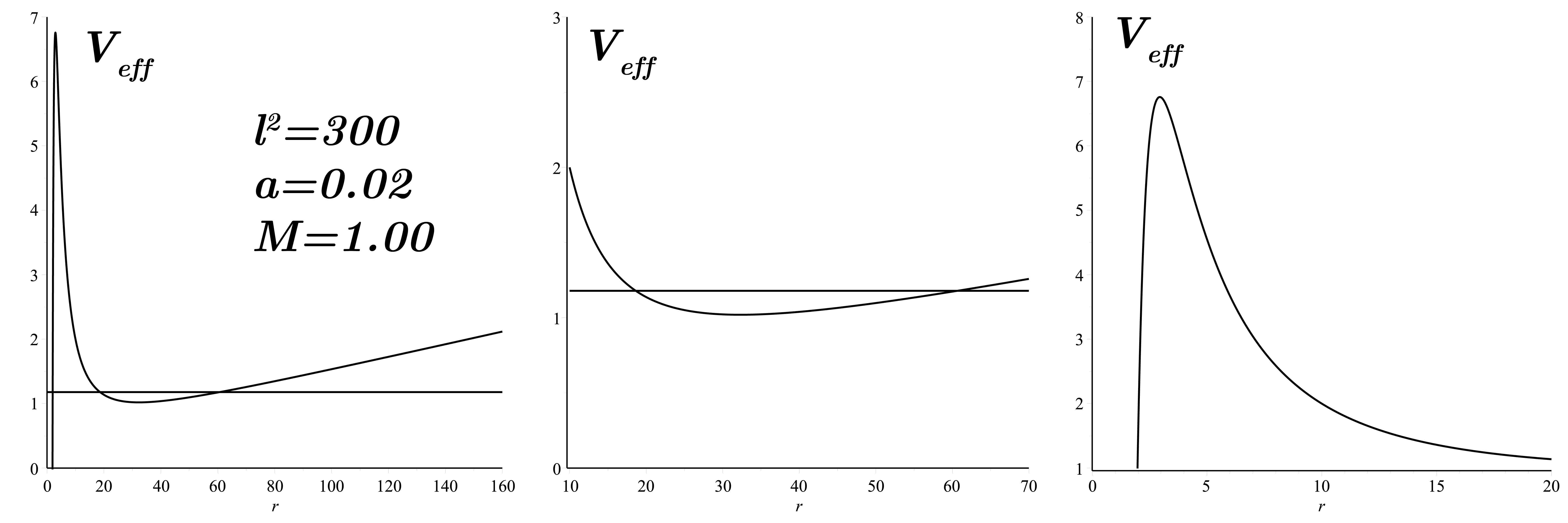}
\caption{$V_{eff}$ for massive particles versus $r$ as a particular example
for $M=1,$ $a=0.01$ and $\ell ^{2}=300.$ The inscriptions depict enlarged
form of unstable and stable parts. The horizontal line shows two turning
points.}
\end{figure}

\subsubsection{Time-like geodesic $\protect\epsilon =1$}

A similar argument is valid also for a massive particle. We set $\epsilon =1$
in Eq.s (35) and (36) to get 
\begin{equation}
\ell ^{2}=\frac{\left( Mu_{c}^{2}+a\right) }{u_{c}^{2}\left(
a+u_{c}-3Mu_{c}^{2}\right) }
\end{equation}%
and 
\begin{equation}
E^{2}=\frac{4\left( a+\frac{u_{c}}{2}-Mu_{c}^{2}\right) ^{2}}{u_{c}\left(
a+u_{c}-3Mu_{c}^{2}\right) }.
\end{equation}%
As we mentioned before here $r_{c}=\frac{1}{u_{c}}$ is the radius of the
equilibrium circular orbit. Fig. 7 shows the variation of $\ell ^{2}/M^{2}$
versus $r/M$ (Eq. (43)) in terms of $a$ which clearly shows that once $%
r\rightarrow r_{c}$ the angular momentum goes to infinity. In the case of $%
\ell ^{2},$ as one can see from Eq. (43) and Fig. 7, $\frac{r}{M}=4$ is the
only orbit in which irrespective to the value of $a$, $\frac{\ell ^{2}}{M^{2}%
}=16.$ 
\begin{figure}[tbp]
\includegraphics[width=180mm,scale=1]{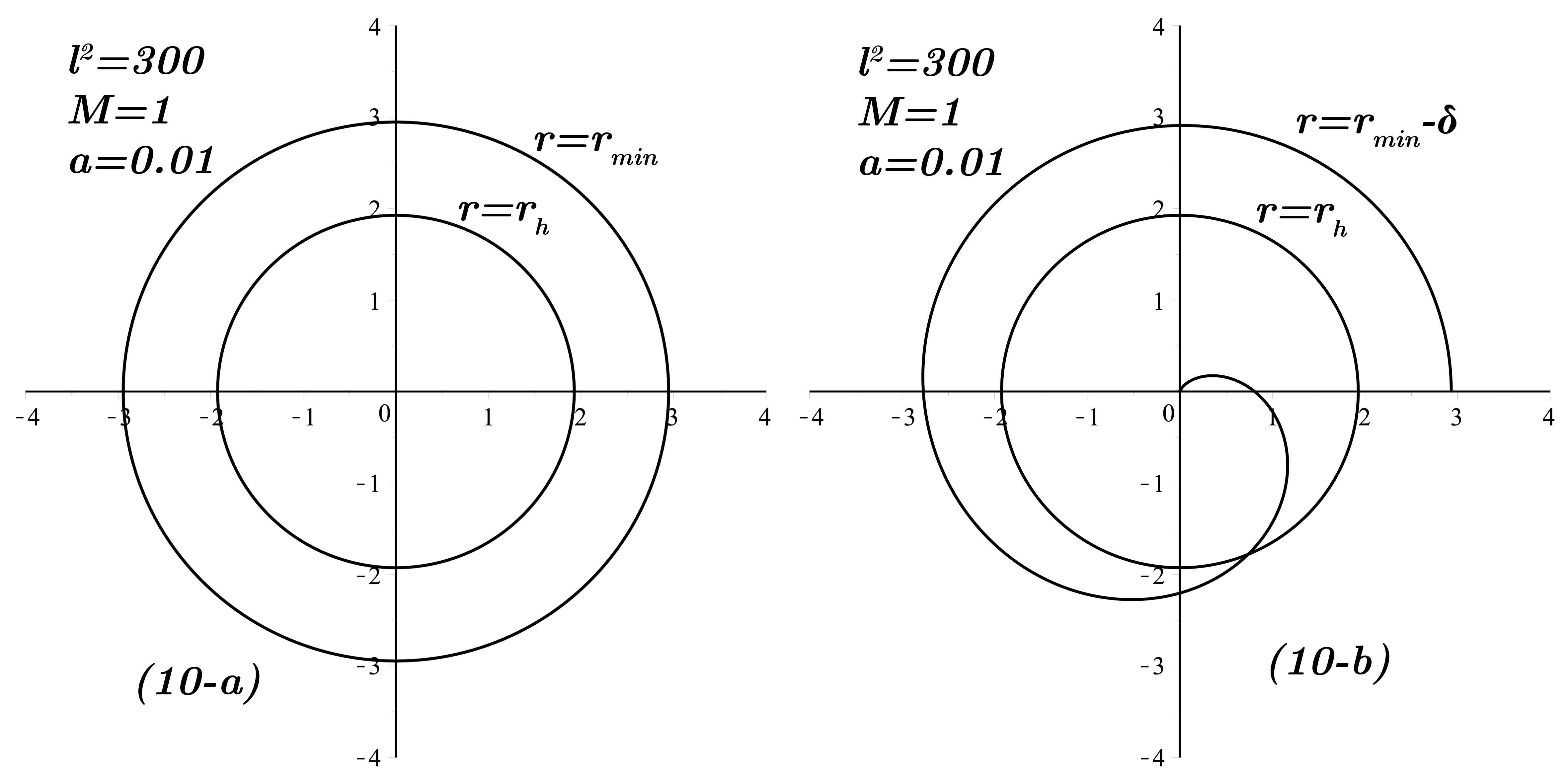}
\caption{a. Details of the circular orbits from Eq. (18), ($\frac{dr}{d%
\protect\phi }=0$). From $E^{2}=13.51647495=(1-\frac{2}{r}+2.02r)\left( 1+%
\frac{300}{r^{2}}\right) $ the minimum circular radius becomes $%
r_{c}=r_{\min }=2.944698040$ whereas the horizon radius is $%
r_{h}=1.925824025.$ b. A small perturbation of $r_{\min }$ in 10a by $%
r_{0}=r_{\min }-\protect\delta $ where $\protect\delta $ is a small positive
number $\left( \protect\delta \ll 1\right) $ causes the geodesics to plunge
into the singularity. This is the fate of the unstable orbits. The
differential equation is solved by "A seventh-eighth order continuous
Runge-Kutta method (dverk78)" method and then the results have been ploted.
The tolerances for the absolute and relative local error are chosen to be $%
10^{-8}.$}
\end{figure}

To investigate the stability of the equilibrium circular motion of a massive
particle we consider the effective potential of the dynamic motion of the
particle given in (16). Again in a stable circular motion the requirements
are $\left. V_{eff}^{\prime }\right\vert _{r_{c}}=0$ and $\left.
V_{eff}^{\prime \prime }\right\vert _{r_{c}}>0$.

Figures 8a and 8b display the effective potential in terms of $r/M$ for $%
a=0.00$ and $a=0.01$ respectively. We see almost similar behavior but still
in the presence of $a$ the stable orbits occur with larger $\ell $ (for a
given mass). The most important difference between the two curves occurs in
the behavior of the graphs to the right of the minimum points. As it is seen
in Fig. 8a, the graph eventually is concave down and becomes asymptotically
constant while in Figure 8b it is concave up and steeply rising. These
indicate that (8b) is different from the $a=0$ case where for a given energy
larger than the asymptotic value of $V_{eff}$ particle would escape to
infinity. In the presence of $a,$ particles can't escape to infinity. Even
if the energy is bigger than the local maximum, to the left of the minimum,
the particle would fall into the singularity of the spacetime.

In Fig. 9 we plot $V_{eff}$ in terms of $r$ for $M=1$, $a=0.01$ and a
particular value for $\ell ^{2}=300.$ The minimum and maximum of the
potential are highlighted. Fig. 10a shows the circular orbit of a massive
particle with unit mass at $r_{c}=$ $r_{\min }$. An infinitesimal deviation
from $r_{\min }$ yields a particle orbit falling into the singularity (Fig.
10b). In Fig. 11a the stable circular orbit at $r_{c}=$ $r_{\max }$ is shown
explicitly while in Fig. 11b the perturbed orbits confined around $r_{\max }$
is depicted. 
\begin{figure}[tbp]
\includegraphics[width=180mm,scale=1]{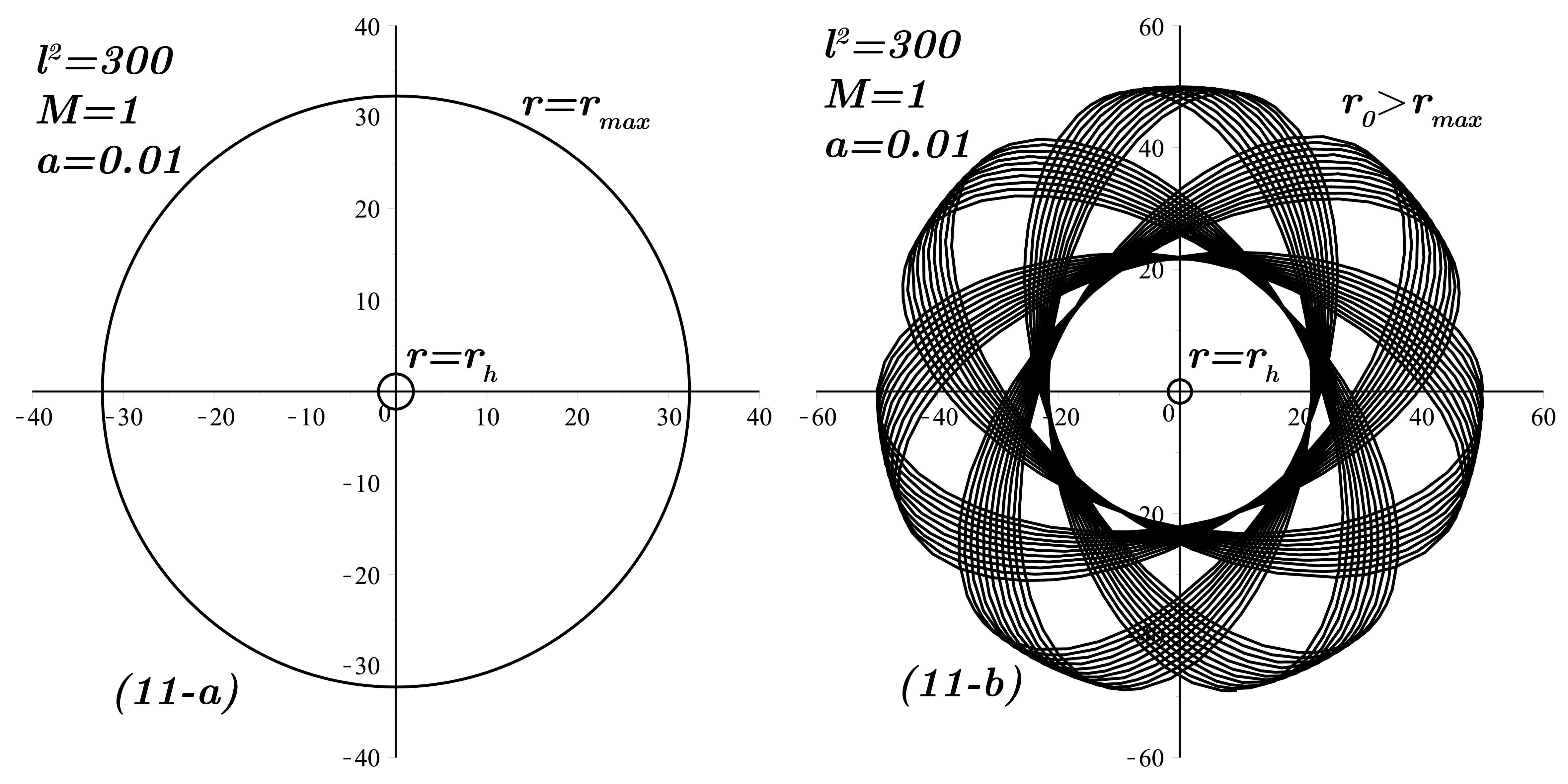}
\caption{The circular orbits displaying $r_{\max }$ at the minimum for the $%
V_{eff}$ in Fig. 9. Any distortion of the particle around $r_{\max }$ in
11-a gives the closed pattern around $r_{\max }$ so that the particle
becomes confined between two radii. Recall that $r_{0}$ is the initial
radial position of the particle in the potential well. The differential
equation is solved by "A seventh-eighth order continuous Runge-Kutta method
(dverk78)" method and then the results have been ploted. The tolerances for
the absolute and relative local error are chosen to be $10^{-8}.$}
\end{figure}

\section{CONCLUSION}

In this paper, the geodesic analysis of the Grumiller's metric which
describes gravity at large distances is considered. The effect of the
Rindler parameter is investigated and compared with the Schwarzschild
geodesics. Timelike and null geodesics are obtained numerically both for
radial and circular motion. Exact radial solutions also are available for
null geodesics as shown in the Appendix. Since our choice is $a>0$, this
amounts to the confinement of geodesics around the central object outside
the event horizon. Closed periodic orbits bounded between two radii are
depicted in Fig. (11b).

For the radial motion; the motion of the particle is bounded with the
inclusion of the Rindler parameter. Hence, unlike the Schwarzschild case,
the particle can not escape to infinity. This is not unexpected since both
mass and Rindler terms are attractive. Plunging into the central singularity
from unstable geodesics due to small perturbation is favored by virtue of
the Rindler acceleration term.

In conclusion, without identifying its physical origin the extra term $\sim
2ar$ in the spherical geometry adds much novelties and richness to the
Schwarzschild geodesics. Given the astrophysical distances in spite of the
small $a\sim 10^{-12}m/s^{2}$, the Rindler term $\sim 2ar$ becomes still
significant. By adjusting a small Rindler parameter flat rotation curves can
be obtained as depicted in Fig. 12. We recall that the velocity $v(r)$ (with
unit mass) is given from the relation $\left\vert Force\right\vert =\frac{M}{%
r^{2}}+a=\frac{v^{2}}{r}$ \cite{1}. 
\begin{figure}[tbp]
\includegraphics[width=180mm,scale=1]{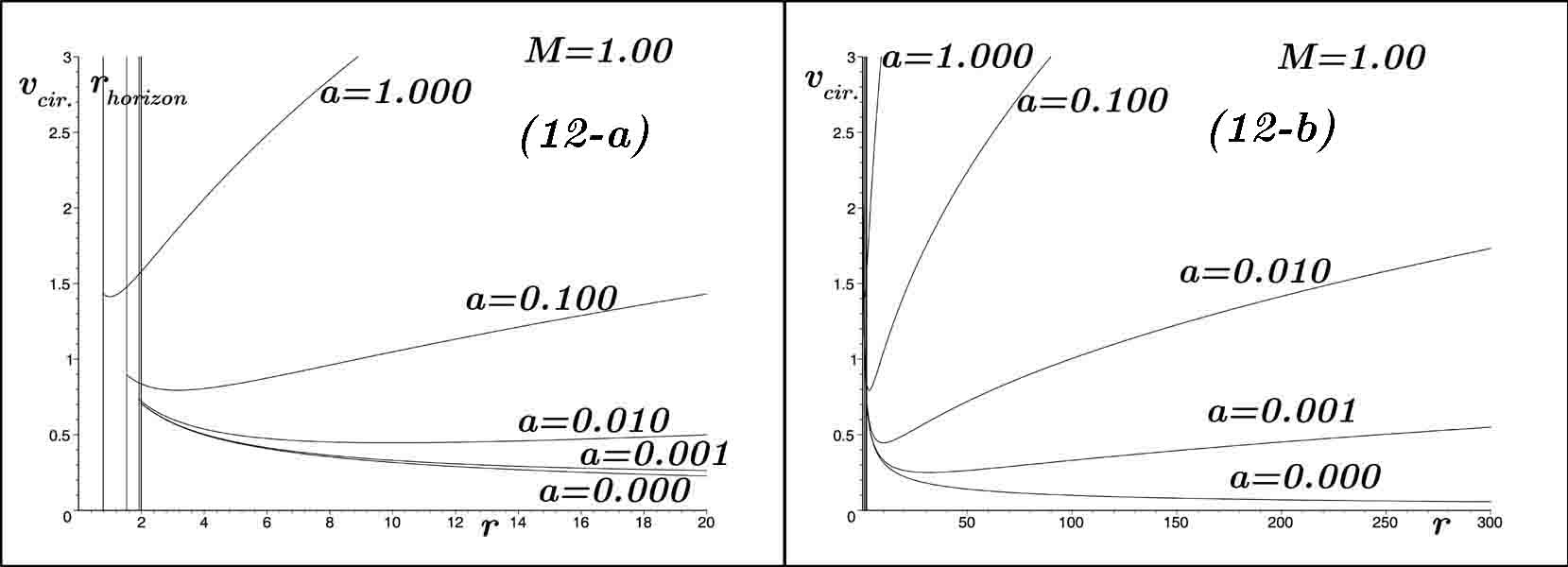}
\caption{Circular velocity plots versus radial distances, at small (a) and
large (b) scales. It is observed that in order to obtain flat rotation
curves i.e. horizontal line $v\left( r\right) $ for circular velocity versus
distance the Rindler acceleration parameter must be rather small (i.e. $a\ll
1$). At the smaller scale (a) the horizon ($r_{h}$) is also shown.}
\end{figure}

\begin{figure}[tbp]
\includegraphics[width=60mm,scale=1]{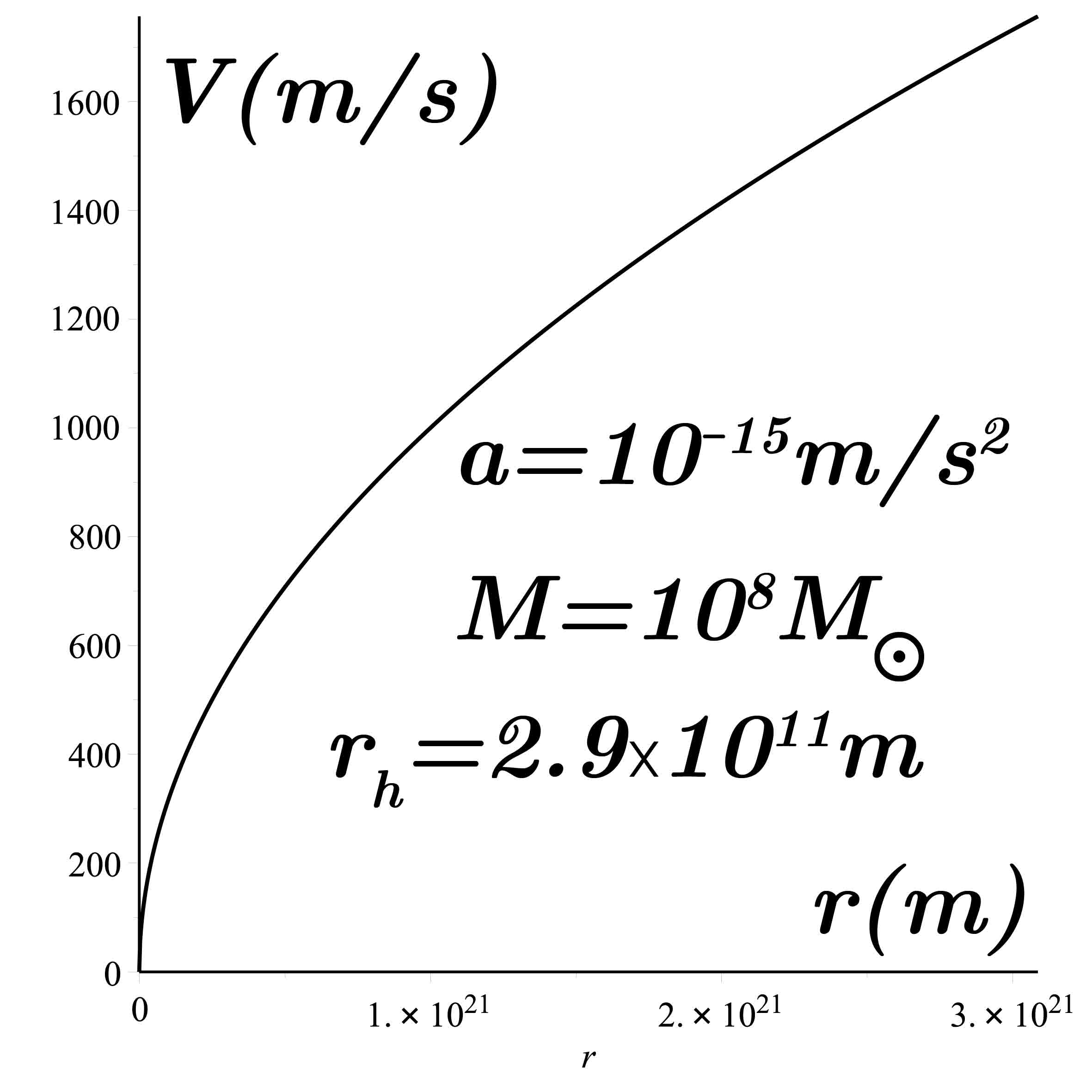}
\caption{Circular velocity plots versus radial distances $r$ for a realistic
value of $a=10^{-15}m/s^{2}$. The value of $M$ is taken from \protect\cite{1}
for a dwarf galaxies i.e., $M=10^{8}M_{\odot }.$ As it is observed the
effect of Rindler acceleration is strong for $r>10^{20}m.$}
\end{figure}

\appendix*

\section{Exact Elliptic Solution to the null Geodesic Equations}

Let's consider the Lagrangian for variable $\theta :$%
\begin{equation}
\mathcal{L}=-\frac{1}{2}f\left( r\right) \left( \frac{dt}{d\lambda }\right)
^{2}+\frac{1}{2f\left( r\right) }\left( \frac{dr}{d\lambda }\right) ^{2}+%
\frac{1}{2}r^{2}\left( \frac{d\theta }{d\lambda }\right) ^{2}+\frac{1}{2}%
r^{2}\sin ^{2}\theta \left( \frac{d\varphi }{d\lambda }\right) ^{2}
\end{equation}%
where $\lambda $ is the affine parameter. Employing the Mino time \cite{10} $%
\gamma $ which is defined as $d\lambda =r^{2}d\gamma $ for convenience it
becomes%
\begin{equation}
\mathcal{L}=-\frac{1}{2}\frac{f\left( r\right) }{r^{4}}\left( \frac{dt}{%
d\gamma }\right) ^{2}+\frac{1}{2r^{4}f\left( r\right) }\left( \frac{dr}{%
d\gamma }\right) ^{2}+\frac{1}{2r^{2}}\left( \frac{d\theta }{d\gamma }%
\right) ^{2}+\frac{1}{2r^{2}}\sin ^{2}\theta \left( \frac{d\varphi }{d\gamma 
}\right) ^{2}
\end{equation}%
which together with the metric condition%
\begin{equation}
-\frac{f\left( r\right) }{r^{4}}\left( \frac{dt}{d\gamma }\right) ^{2}+\frac{%
1}{r^{4}f\left( r\right) }\left( \frac{dr}{d\gamma }\right) ^{2}+\frac{1}{%
r^{2}}\left( \frac{d\theta }{d\gamma }\right) ^{2}+\frac{1}{r^{2}}\sin
^{2}\theta \left( \frac{d\varphi }{d\gamma }\right) ^{2}=\epsilon
\end{equation}%
yield the geodesics equations as:%
\begin{equation}
\frac{dt}{d\gamma }=\frac{r^{3}}{\Delta }\alpha
\end{equation}%
\begin{equation}
\frac{d\varphi }{d\gamma }=\frac{\beta }{\sin ^{2}\theta }
\end{equation}%
\begin{equation}
\left( \frac{d\theta }{d\gamma }\right) ^{2}=k-\frac{\beta ^{2}}{\sin
^{2}\theta }
\end{equation}%
and 
\begin{equation}
\left( \frac{dr}{d\gamma }\right) ^{2}=\left( \epsilon r^{2}-k\right)
r\Delta +r^{4}\alpha ^{2}.
\end{equation}

\subsection{The $r$-equation}

In these equations $k,$ $\alpha $ and $\beta $ are integration constants and 
$\Delta =rf\left( r\right) =r-2M+2ar^{2}.$ Following the method introduced
in \cite{11} $r$-equation (A7), after introducing $r=\pm \frac{1}{x}+r_{0}$
in which $r_{0}$ is a root for $\left( \epsilon r^{2}-k\right) r\Delta
+r^{4}\alpha ^{2}=0$ and setting $\epsilon =0$ becomes%
\begin{equation}
\left( \frac{dx}{d\gamma }\right) ^{2}=b_{0}+b_{1}x+b_{2}x^{2}+b_{3}x^{3}
\end{equation}%
in which%
\begin{eqnarray}
b_{0} &=&\alpha ^{2} \\
b_{1} &=&\pm \frac{2r_{0}\alpha ^{2}\left( 2r_{0}-4M+3ar_{0}^{2}\right) }{%
r_{0}-2M+2ar_{0}^{2}}  \notag \\
b_{2} &=&\frac{r_{0}^{2}\alpha ^{2}\left( 5r_{0}-12M+6ar_{0}^{2}\right) }{%
r_{0}-2M+2ar_{0}^{2}}  \notag \\
b_{3} &=&\pm \frac{2r_{0}^{3}\alpha ^{2}\left( r_{0}-3M+ar_{0}^{2}\right) }{%
r_{0}-2M+2ar_{0}^{2}}.  \notag
\end{eqnarray}%
Finally a further transformation 
\begin{equation}
x=\frac{4y-\frac{b_{2}}{3}}{b_{3}}
\end{equation}%
eliminates the second order and (A8) reads 
\begin{equation}
\left( \frac{dy}{d\gamma }\right) ^{2}=4y^{3}-g_{2}y-g_{3}
\end{equation}%
where%
\begin{equation}
g_{2}=\frac{b_{2}^{2}-3b_{1}b_{3}}{12}
\end{equation}%
and%
\begin{equation}
g_{3}=\frac{b_{1}b_{2}b_{3}}{48}-\frac{b_{2}^{3}}{216}-\frac{b_{0}b_{3}^{2}}{%
16}.
\end{equation}%
The final form of the equation (A11) is nothing but of elliptic type and its
solution is the WeierstrassP function \cite{12} i.e, 
\begin{equation}
y\left( \gamma \right) =WeierstrassP\left( \gamma -\gamma
_{0},g_{2},g_{3}\right)
\end{equation}%
where $\gamma _{0}$ is an integration constant. One obtains as a result%
\begin{equation}
r=\pm \frac{b_{3}}{4WeierstrassP\left( \gamma -\gamma
_{0},g_{2},g_{3}\right) -\frac{b_{2}}{3}}+r_{0}.
\end{equation}%
Let us note that for $\epsilon =1$ i.e. for time like geodesics, the
foregoing method does not work, at least the solution is not in elliptic
form. For this reason we considered the null geodesics alone.

\subsection{$\protect\theta $ and $\protect\varphi $-equations}

$\theta $-equation (A6) can be easily solved and the analytic answer is
given by%
\begin{equation}
\theta _{\pm }\left( \gamma \right) =\pi \pm \cos ^{-1}\left( \sqrt{\beta }%
\left( \gamma -\tilde{\gamma}_{0}\right) \right)
\end{equation}%
in which $\tilde{\gamma}_{0}$ is an integration constant. After $\theta $
one can integrate the $\varphi $-equation to find%
\begin{equation}
\varphi \left( \gamma \right) =\sqrt{\beta }\left( \sqrt{\beta }\left(
\gamma -\tilde{\gamma}_{0}\right) \right) +\varphi _{0}
\end{equation}%
in which $\varphi _{0}$ is an integration constant.

\bigskip

\bigskip

\end{document}